\documentclass[preprint,floatfix] {revtex4} 
\newcommand{\rvec}{\mathrm {\mathbf {r}}} 
\newcommand{\pvec}{\mathrm {\mathbf {p}}} 
\usepackage{graphicx}
\usepackage{subfigure}
\usepackage{xcolor}
\usepackage{amsmath}
\usepackage{enumerate}
\usepackage{longtable}
\usepackage{tabularx}
\usepackage{amssymb}

\usepackage{color, soul}
\definecolor{darkblue}{rgb}{0,0,0.5}
\setulcolor{darkblue}

\begin{document}

\title{Various complexity measures in confined hydrogen atom}

\author{Sangita Majumdar}
\author{Neetik Mukherjee}

\author{Amlan K.~Roy}
\altaffiliation{Corresponding author. Email: akroy@iiserkol.ac.in, akroy6k@gmail.com.}
\affiliation{Department of Chemical Sciences\\
Indian Institute of Science Education and Research (IISER) Kolkata, 
Mohanpur-741246, Nadia, WB, India}

\begin{abstract}

Several well-known statistical measures similar to \emph{LMC} and \emph{Fisher-Shannon} complexity have been computed for confined 
hydrogen atom in both position ($r$) and momentum ($p$) spaces. Further, a more generalized form of these quantities 
with R\'enyi entropy ($R$) is explored here. The role of scaling parameter in the exponential part is also pursued. $R$ is 
evaluated taking order of entropic moments $\alpha, \beta$ as $(\frac{2}{3},3)$ in $r$ and $p$ spaces. Detailed systematic 
results of these measures with respect to variation of confinement radius $r_c$ is presented for low-lying states such as, 
$1s$-$3d,~4f$ and $5g$. For \emph{nodal} states, such as $2s,~3s$ and $3p$, as $r_c$ progresses there appears a maximum followed 
by a minimum in $r$ space, having certain values of the scaling parameter. However, the corresponding $p$-space results lack such 
distinct patterns. This study reveals many other interesting features. 

\vspace{5mm}

\vspace{5mm}

\end{abstract}
\maketitle

\section{introduction}
Quantum particles experience intense changes in their physical and chemical properties under spatial confinement. Spherically 
confined quantum systems have been explored extensively \cite{sabin2009}, with wide-spread applications in quantum dot, quantum 
wells and quantum wires, etc. Dramatic changes in various observable properties such as energy spectrum, transition frequencies, 
transition probabilities, polarizability, chemical reactivity, ionization potential etc., were reported to occur under 
such situations \cite{sabin2009,sen2014electronic}.

Recently there has been a growing interest in using statistical quantities namely, Fisher information ($I$), Onicescu energy ($E$), 
Shannon entropy ($S$) and R\'enyi entropy ($R$) as descriptors of certain chemical, physical properties of a quantum system. Along 
these lines, \emph{complexity}, another relevant concept, is directly related to aforementioned 
measures, representing their combined effect. A universal characterization has not been possible, but can be proposed as 
an indicator of pattern, structure or correlation associated with the distribution function in a given system. It depends on 
the scale of observation, and constitutes an important area of research with contemporary interest in disordered systems, spatial 
patterns, language, multi-electronic systems, molecular or DNA analysis, social science, 
\cite{rosso03,shalizi04,chatzisavvas05,bouvrie11} etc.  

An atom is a complex system; restricting its motion in an enclosure makes it even more fascinating according to a complex 
world \cite{goldenfeld99,sen12}. Complexity, in a system, arises due to breakdown of certain symmetry rules. For finite complexity, 
the system is either in a state having some less than maximal $order$ or not in equilibrium. Stated differently, it vanishes at 
two limiting cases, \emph{viz.}, when it is (a) at equilibrium (maximum disorder) or b) completely ordered (maximum distance from 
equilibrium) \cite{sen02,sen12}. It gives a qualitative idea of organization in a system and is considered as a 
general indicator of structure and correlation. In literature various definitions are available; some of them are 
Shiner, Davidson, Landsberg (\emph{SDL}) \cite{landsberg84,landsberg98,shiner99}, L\'opez-Ruiz, Mancini, Calbet 
(LMC) \emph{shape} ($C_{LMC}$) \cite{lopezruiz95,anteneodo96,catalan02,sanchez05}, \emph{Fisher-Shannon} 
($C_{IS}$) \cite{sen07,angulo08}, \emph{Cram\'er-Rao} \cite{dehesa06,angulo08,antolin09} or 
\emph{Generalized R\'enyi-like} complexity \cite{calbet01,martin06,romera08}, etc.  

The statistical measure of complexity, in product form, can be written as,
\begin{equation}
 C_{LMC} = H . D
\end{equation}
where \emph{H} represents the information content and \emph{D} gives an idea of concentration of spatial distribution. 
In order to satisfy conditions such as reaching minimal values for both extremely ordered and disordered limits, invariance 
under scaling, translation and replication, this quantity was criticized \cite{feldman98} and modified \cite{lopez05},  
giving rise to the expression, 
\begin{equation}
C_{\emph{LMC}} = D . {e^S}.
\end{equation}
Principally this gives an interplay between information stored in a system, and measure of a probabilistic 
hierarchy amongst its observed parts. It has application in diverse fields like detection of periodic, 
quasi-periodic, linear stochastic, chaotic dynamics \cite{lopezruiz95,yamano04,yamano04a}.

In information theory $E$ signifies a measure of order, because it becomes minimum at equilibrium. Whereas, information 
entropies like $S, R$, being maximum at equilibrium, signify disorder. Complexity identifies the extent of balance between 
order and disorder. Sometimes, $I$ is used in place of $E$. So far $S$ has been extensively used as disorder parameter. 
$C_{IS}$ is another measure, obtained by replacing the pre-exponential global factor in $C_{LMC}$ by a local factor 
like $I$. It combines global and local characters while preserving desirable properties of complexity. Usefulness
of $C_{IS}$ can be judged by looking at the numerous works done for both \emph{free} and \emph{confined} atomic systems, like 
atomic shell structure, ionization process \cite{angulo08,sen07,antolin09,angulo08a} etc. A more generalized version was also 
proposed that uses $R$ in place of $S$, in $C_{LMC}$ and $C_{IS}$ \cite{angulo12}. Later, a scaling factor ($b$) was introduced 
in exponential part. 

About a decade ago, $C_{LMC}$ was used in the context of Rydberg states of \emph{free} hydrogen atom (FHA) in $r, p$ spaces 
\cite{sanudo08}. Later, $C_{LMC}$ and SDL complexity was employed in atoms \cite{sen12}. However, in a \emph{confined} hydrogen 
atom (CHA) complexity measures have been pursued rather rarely. Two major works in this direction involved calculation of 
$C_{IS}$ in composite space for ground state of CHA under soft and hard confinement \cite{aquino13,nagy09}. In this endeavor, 
our focus is to explore four different types of complexity arising out of two order ($I, E$) and two disorder ($S, R$) parameters, 
in both space as functions of confinement radius ($r_c$). As in the literature \cite{sen12}, we also adopt two $b$ values 
($\frac{2}{3}$ for $C_{IS}$, 1 for $C_{LMC}$). All calculations were done using the \emph{exact} wave functions of 
CHA in $r$ space. The $p$-space wave function is obtained from numerical Fourier transform of $r$-space counterpart. In the end, 
pilot calculation are done for eight low-lying states \emph{viz.,} $1s$-$3d, 4f, 5g$. Organization of this article is as follows. 
Section~II gives a brief account of the theoretical method used; Sec.~III presents a detailed discussion on our results, while 
we conclude with a few remarks in Sec.~IV.   

\section {Methodology}
Exact radial wave function for a CHA can be expressed as \cite{burrows06},
\begin{equation}
\psi_{n, l}(r)= N_{n, l}\left(2r\sqrt{-2\mathcal{E}_{n,l}}\right)^{l} \ _{1}F_{1}
\left[\left(l+1-\frac{1}{\sqrt{-2\mathcal{E}_{n,l}}}\right),(2l+2),2r\sqrt{-2\mathcal{E}_{n,l}}\right] 
e^{-r\sqrt{-2\mathcal{E}_{n,l}}}.  
\end{equation}
Here, $N_{n, l}$ denotes normalization constant and $\mathcal{E}_{n,l}$ corresponds to energy of a given state characterized by 
radial and angular quantum numbers $n,l$ respectively, whereas $_1F_1\left[a,b,r\right]$ represents confluent hypergeometric 
function. Allowed energies are obtained by imposing the boundary condition $\psi_{n,\ell} (0)= \psi_{n,\ell} \ (r_c)=0$. In this 
work, generalized pseudospectral (GPS) method was employed to estimate $\mathcal{E}_{n,l}$ of these states. This method has 
provided very accurate results for various model and real systems including atoms, molecules, some of which could be found in the 
references \cite{roy04,sen06,roy15}. 
 
The $p$-space wave function is obtained from Fourier transform of $r$-space counterpart,  
\begin{equation}
\begin{aligned}
\psi_{n,l}(p) & = & \frac{1}{(2\pi)^{\frac{3}{2}}} \  \int_0^\infty \int_0^\pi \int_0^{2\pi} \psi_{n,l}(r) \ \Theta(\theta) 
 \Phi(\phi) \ e^{ipr \cos \theta}  r^2 \sin \theta \ \mathrm{d}r \mathrm{d} \theta \mathrm{d} \phi   \\
      & = & \frac{1}{2\pi} \sqrt{\frac{2l+1}{2}} \int_0^\infty \int_0^\pi \psi_{n,l} (r) \  P_{l}^{0}(\cos \theta) \ 
e^{ipr \cos \theta} \ r^2 \sin \theta  \ \mathrm{d}r \mathrm{d} \theta.  
\end{aligned}
\end{equation}
Here $\psi(p)$ is not normalized and needs to be normalized. Integrating over $\theta$ and $\phi$ yields,  
\begin{equation}
\psi_{n,l}(p)=(-i)^{l} \int_0^\infty \  \frac{\psi_{n,l}(r)}{p} \ f(r,p)\mathrm{d}r.    
\end{equation}
Depending on $l$, this can be rewritten in following simplified form ($m$ starts with 0), 
\begin{equation}
\begin{aligned}
f(r,p) & = & \sum_{k=2m+1}^{m<\frac{l}{2}} a_{k} \ \frac{\cos pr}{p^{k}r^{k-1}} +  
\sum_{j=2m}^{m=\frac{l}{2}} b_{j} \ \frac{\sin pr}{p^{j}r^{j-1}}, \ \ \ \ \mathrm{for} \ \mathrm{even} \ l,   
\\ f(r,p) & = & \sum_{k=2m}^{m=\frac{l-1}{2}} a_{k} \ \frac{\cos pr}{p^{k}r^{k-1}} +  
\sum_{j=2m+1}^{m=\frac{l-1}{2}} b_{j} \ \frac{\sin pr}{p^{j}r^{j-1}}, \ \ \ \ \mathrm{for} \ \mathrm{odd} \ l.
\end{aligned} 
\end{equation}
The values of coefficients $a_{k}$, $b_{j}$ of even-$l$ and odd-$l$ states can easily be computed from Eq.~(2).
Normalized position and momentum electron densities are expressed as,
\begin{equation}
\begin{aligned}
\rho(\rvec) = |\psi_{n,l,m}(\rvec)|^2 ,    \ \ \ \ \ \ 
\Pi(\pvec) = |\psi_{n,l,m} (\pvec)|^2 .
\end{aligned}
\end{equation}

Without any loss of generality, let us define complexity in following general form $C = Ae^{b.B}$. The order ($A$) and disorder 
parameters ($B$) may include ($E, I$) and ($R, S$) respectively. With this in mind, we are interested in the following four
quantities, 
\begin{equation}
\begin{aligned}
C_{ER} & = E e^{bR}, \ \ \ \ \ \ \ \ C_{IR} = Ie^{bR}, \ \ \ \ \ \ \ C_{ES} & = E e^{bS}, \ \ \ \ \ \ \ \ C_{IS} = Ie^{bS}. 
\end{aligned} 
\end{equation}

Shannon entropies of a continuous density distribution are written as (`t' stands for \emph{total}), 
\begin{equation}
\begin{aligned} 
S_{\rvec} & =  -\int_{{\mathcal{R}}^3} \rho(\rvec) \ \ln [\rho(\rvec)] \ \mathrm{d} \rvec ;\ \ \ 
S_{\pvec} & =  -\int_{{\mathcal{R}}^3} \Pi(\pvec) \ \ln [\Pi(\pvec)] \ \mathrm{d} \pvec; \ \ \
S_{t} & = S_{\rvec}+S_{\pvec}. 
\end {aligned}
\end{equation}
Similarly, R{\'e}nyi entropies of order $\alpha (\neq 1)$ are obtained by taking logarithm of $\alpha$ and $\beta$-order 
entropic moments in respective spaces,  
\begin{equation}
\begin{aligned}
R_{\rvec}^{\alpha}  =  \frac{1}{1-\alpha} \ln \left(\int_{{\mathcal{R}}^3} \rho^{\alpha}(\rvec)\mathrm{d} \rvec \right) ;\ \ \ 
R_{\pvec}^{\beta}  =  \frac{1}{1-\beta} \ln \left[\int_{{\mathcal{R}}^3} \Pi^{\beta}(\pvec)\mathrm{d} \pvec \right]; \ \ \
R_{t} = R_{\rvec}^{\alpha}+R_{\pvec}^{\beta}. 
\end {aligned}
\end{equation}
The general form of $I_{\rvec}$, $I_{\pvec}$ for a particle in a central potential may be simplified as \cite{romera05},
\begin{equation}
\begin{aligned} 
I_{\rvec} & =  4\langle p^2\rangle - 2(2l+1)|m|\langle r^{-2}\rangle;\ \ \
I_{\pvec} & =  4\langle r^2\rangle - 2(2l+1)|m|\langle p^{-2}\rangle; \ \ \
I_{t} = I_{\rvec}I_{\pvec}.  
\end{aligned} 
\end{equation}
Finally, $E$ is given by the following expressions in conjugate space, 
\begin{equation}
\begin{aligned}
E_{\rvec} & = \int_{{\mathcal{R}}^3}\rho^{2}(\rvec)\mathrm{d} \rvec;\ \ \ 
E_{\pvec} & = \int_{{\mathcal{R}}^3}\Pi^{2}(\pvec)\mathrm{d} \pvec;\ \ \
E_{t} = E_{\rvec}E_{\pvec}.  
\end{aligned}
\end{equation}

\section{Result and Discussion} 
At first let us clear a few things before we begin our discussion. All the tables and figures that follow quote the \emph{net} 
information measures in conjugate $r$ and $p$ space of CHA, which may be partitioned in to radial and angular contributions.
In a given space, all results correspond to \emph{net} measures including the \emph{angular} parts.
By squeezing the radial boundary of FHA from infinity to a finite region, one progresses to a CHA. As this does not alter the
\emph{angular} boundary conditions, angular portion of the information measures in FHA and CHA remains unchanged in both 
spaces. Further as we are solely interested in \emph{radial} confinement, same will also not change as one modifies $r_c$
values from one to another. However, there will be non-vanishing contribution from $l, m$ quantum numbers. Throughout our 
calculation, magnetic quantum number $m$ is set to $0$. All the aforementioned measures of Eq.~(8) have been investigated 
with respect to $r_c$, for two selected values of $b$ ($1,\frac{2}{3}$); these are the ones which are widely used in literature.   
Note that, for $b=1$, $C_{ES}^{(2)}$ reduces to $C_{LMC}$; similarly $C_{IS}^{(1)}$ corresponding to $b=\frac{2}{3}$ 
refers to $C_{IS}$ of literature. In order to 
facilitate the discussion, a few words may be devoted to the notation followed. A uniform symbol $C_{order_{s}, disorder_{s}}^b$
is used; where the two subscripts refer to two order ($E, I$) and disorder ($S, R$) parameters. Another subscript $s$ is used
to specify the space; \emph{viz.}, $r, p$ or $t$ (total). Two scaling parameters $b=\frac{2}{3}, 1$ are identified with 
superscripts 1, 2. These measures are offered systematically for $1s$-$3d$ as well as $4f$ and $5g$ states in conjugate spaces, 
with $r_c$ varying in the range of 0.1-100 a.u.  

\begin{figure}                         
\begin{minipage}[c]{0.30\textwidth}\centering
\includegraphics[scale=0.45]{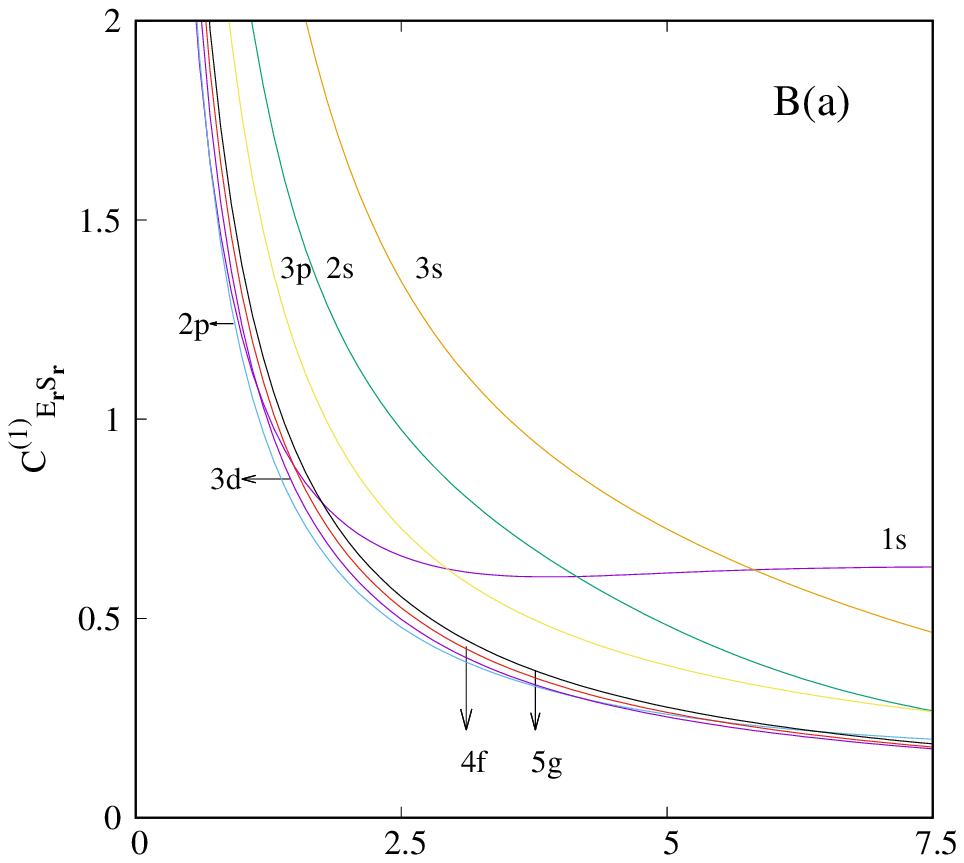}
\end{minipage}%
\vspace{1mm}
\begin{minipage}[c]{0.30\textwidth}\centering
\includegraphics[scale=0.45]{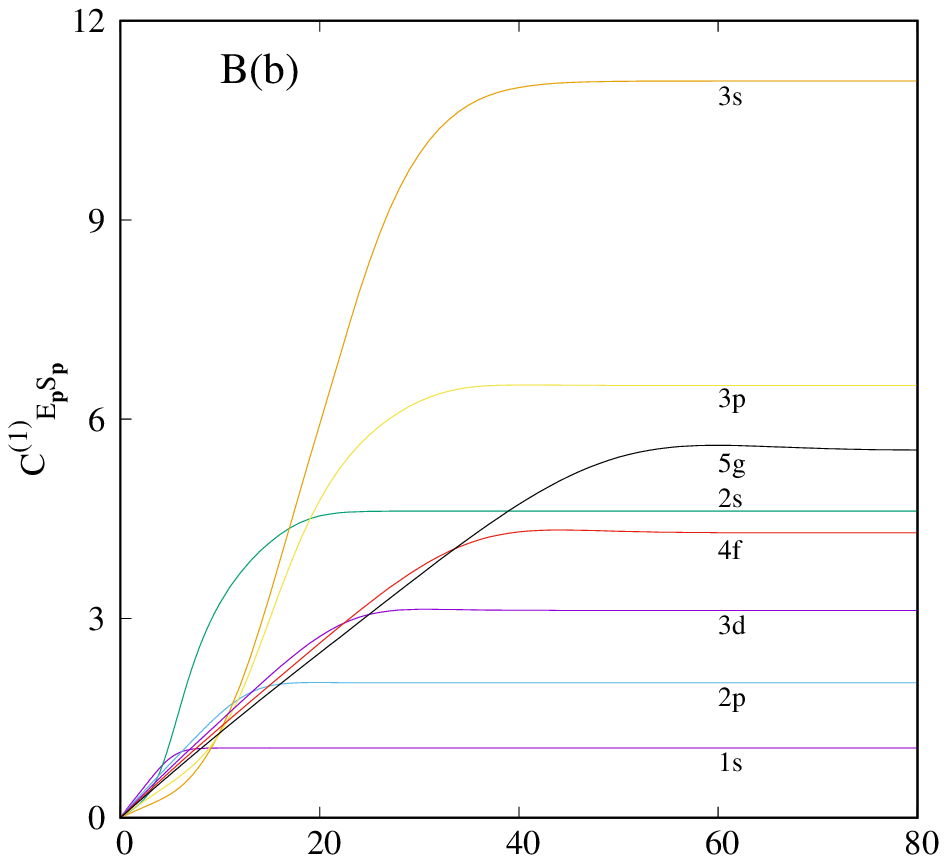}
\end{minipage}%
\vspace{1mm}
\hspace{0.2in}
\begin{minipage}[c]{0.32\textwidth}\centering
\includegraphics[scale=0.46]{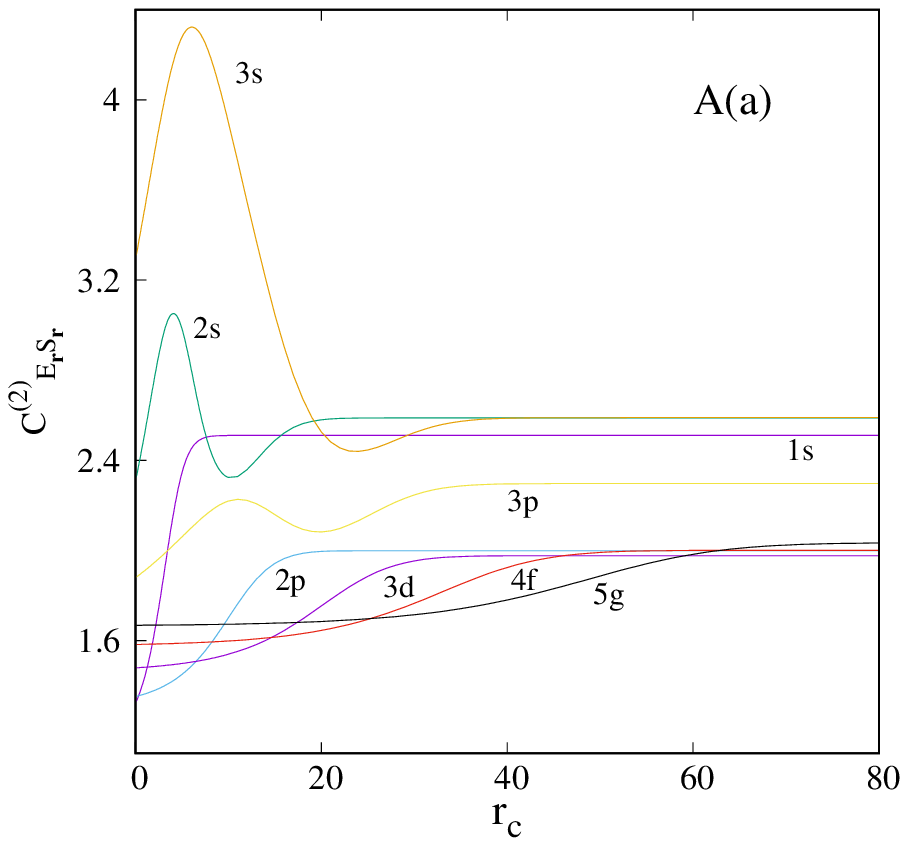}
\end{minipage}%
\begin{minipage}[c]{0.32\textwidth}\centering
\includegraphics[scale=0.46]{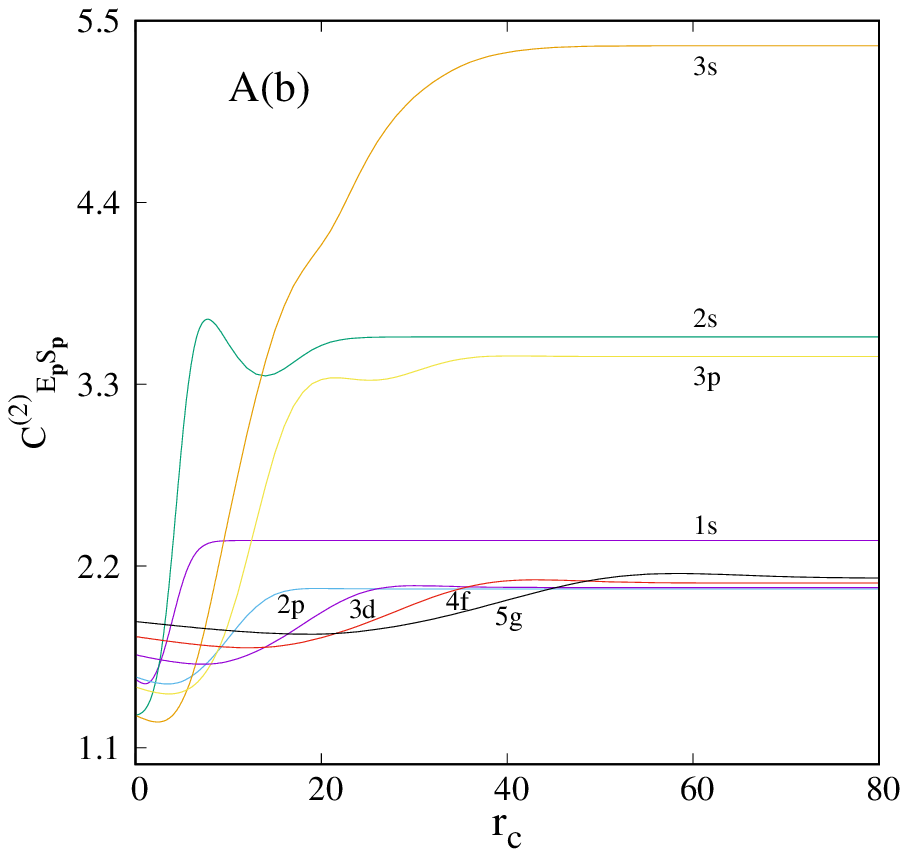}
\end{minipage}%
\caption{Variation of $C_{E_{\rvec}S_{\rvec}}^{(2)},~C_{E_{\pvec}S_{\pvec}}^{(2)}$ (bottom row A) and 
$C_{E_{\rvec}S_{\rvec}}^{(1)},~C_{E_{\pvec}S_{\pvec}}^{(1)}$ (top row B) in CHA with $r_c$ for $1s$-$3d, 4f$ and $5g$
states.  See text for details.}
\end{figure}
 
\begingroup           
\squeezetable
\begin{table}
\caption{$C_{E_{\rvec}S_{\rvec}}^{(2)},~C_{E_{\pvec}S_{\pvec}}^{(2)}$ and $C_{E_{t}S_{t}}^{(2)}$ for $1s,~2s,~2p,~3d$ 
states in CHA at various $r_c$.}
\centering
\begin{ruledtabular}
\begin{tabular}{l|lll|l|lll}
$r_c$  &  $C_{E_{\rvec}S_{\rvec}}^{(2)}$ & $C_{E_{\pvec}S_{\pvec}}^{(2)}$  & $C_{E_{t}S_{t}}^{(2)}$\hspace{10mm} & $r_c$ 
    & $C_{E_{\rvec}S_{\rvec}}^{(2)}$ & $C_{E_{\pvec}S_{\pvec}}^{(2)}$  & $C_{E_{t}S_{t}}^{(2)}$ \vspace{1mm}  \\ 
\hline
\multicolumn{4}{c}{$1s$} \vline &  \multicolumn{4}{c}{\hspace{-5mm}$2s$}    \\
\hline
 0.1   &  1.330123  &   1.5122  &   2.0114  &   0.1 &   2.325609   &  1.2984  &   3.0196  \\
 2.0   &  1.624390  &   1.5304  &   2.4860  &   1.5 &   2.636436   &  1.3766  &   3.6295  \\
 3.5   &  2.012256  &   1.7709  &   3.5636  &   4.1 &   3.051717   &  2.3753  &   7.2489  \\
 4.5   &  2.254326  &   1.9863  &   4.4778  &   6.9 &   2.624591   &  3.6385  &   9.5496  \\
 7.5   &  2.501324  &   2.3284  &   5.8242  &  10.0 &   2.324233   &  3.5457  &   8.2410  \\
10.0   &  2.510443  &   2.3533  &   5.9078  &  13.0 &   2.401653   &  3.3610  &   8.0721  \\
16.0   &  2.510692  &   2.3543  &   5.9110  &  38.0 &   2.588344   &  3.5864  &   9.2830  \\
20.0   &  2.510692  &   2.3543  &   5.9110  &  40.0 &   2.588344   &  3.5864  &   9.2830  \\
\hline
\multicolumn{4}{c}{$2p$}  \vline  &   \multicolumn{4}{c}{\hspace{-5mm}$3d$}    \\
\hline
 0.1  &  1.352613 & 1.5276   & 2.0662  &    0.1      & 1.479327  &   1.6617     &  2.4582   \\
 2.5  &  1.386951 & 1.4907   & 2.0676  &   10.0      & 1.539579  &   1.6231     &  2.4989   \\
 5.0  &  1.449780 & 1.5013   & 2.1766  &   16.0      & 1.644781  &   1.7746     &  2.9189   \\
 8.5  &  1.609479 & 1.6609   & 2.6733  &   20.0      & 1.753859  &   1.9185     &  3.3648   \\
12.0  &  1.821269 & 1.8978   & 3.4565  &   22.0      & 1.811337  &   1.9824     &  3.5909   \\
14.0  &  1.912716 & 1.9966   & 3.8189  &   26.0      & 1.904810  &   2.0619     &  3.9276   \\
33.0  &  1.999251 & 2.0613   & 4.1211  &   56.0      & 1.977128  &   2.0698     &  4.0923   \\
40.0  &  1.999251 & 2.0613   & 4.1211  &   60.0      & 1.977128  &   2.0698     &  4.0923   \\
\end{tabular}
\end{ruledtabular}
\end{table}
\endgroup   

At first, in Fig.~1 $C_{E_{\rvec}S_{\rvec}}^{(2)},~C_{E_{\pvec}S_{\pvec}}^{(2)}$ are plotted against $r_c$ in two bottom panels
A(a) and A(b); two similar plots for $C_{E_{\rvec}S_{\rvec}}^{(1)},~C_{E_{\pvec}S_{\pvec}}^{(1)}$ in top two segments B(a) and
B(b). Note that the range of $r_c$ is same for all four panels except B(a). Panel~A(a) clearly reveals that, 
$C_{E_{\rvec}S_{\rvec}}^{(2)}$ for circular states $(1s,~2p,~3d,~4f,~5g)$ gradually increases with rise of $r_c$ before reaching
a threshold corresponding to the FHA result. The particular $r_c$ at which this limiting value is reached tends to grow as $l$ 
goes up. On the other hand, the same for nodal states ($2s,~3s,~3p$) shows a maximum followed by a minimum with $r_c$ and 
finally converges to respective FHA value. Appearance of such extrema in $C_{E_{\rvec}S_{\rvec}}^{(2)}$ thus can be considered 
as an indication of presence of nodes. Importantly, however, an increase in number of nodes in wave function apparently does not 
affect the number of extrema produced in $C_{E_{\rvec}S_{\rvec}}^{(2)}$. In strong confinement region $(r_{c} \lessapprox 0.4)$, 
for a particular $r_{c}$, $C_{E_{\rvec}S_{\rvec}}^{(2)}$ enhances with $n$ for the circular states. But in the higher $r_{c}$ 
region, significant crossing occurs amongst these states; so this ordering is no longer maintained. On the other hand, at a fixed 
low $r_c$, $C_{E_{\rvec}S_{\rvec}}^{(2)}$ for $2s,~3s,~3p$ states accelerate with number of nodes. When they have equal number of 
nodes then the state with lower $n$ has greater complexity. But at weak confinement region this ordering dissolves. Now, from 
panel~A(b) one infers that, for all these reported states there occur a minimum in $C_{E_{\pvec}S_{\pvec}}^{(2)}$. Position of 
this minimum shifts toward right with increase in both $n$ and $l$. After the minimum point, $C_{E_{\pvec}S_{\pvec}}^{(2)}$ for 
node-less states grows up to reach their terminal value. Whereas for states having nodes ($2s, 3s, 3p$), the minimum is preceded
by a maximum with increment of $r_c$. But importantly, these extrema get flattened with progress of $n$ and fall of $l$ for 
a given $n$. Now, 
panel~A(c) in Fig.~S1 presents variations of the total quantities $C_{E_{t}S_{t}}^{(2)}$ for the concerned states. As usual 
the nodeless states $1s,~2p,~3d,~4f,~5g$ continually increase with $r_c$ until converging to FHA limit, whereas nodal
states go through some extrema before reaching that limit--qualitatively much similar to a pattern encountered in A(a). 
Next, panels B(a) and B(b) in the top row, delineate that, for all these states $C_{E_{\rvec}S_{\rvec}}^{(1)}$ monotonically 
decline and $C_{E_{\pvec}S_{\pvec}}^{(1)}$ enhance with rise of $r_c$ respectively. However, panel B(c) in Fig.~S1 shows that, 
$C_{E_{t}S_{t}}^{(1)}$ for circular states elevate with $r_c$. But, for for the nodal states ($2s,~3s,~3p$), there occur some 
extrema, which thinning out with progress in $n,l$ quantum numbers. From the study of these two sets of complexity measures, 
namely, $C^{(2)}_{ES}$ in A(a)-A(c) and $C^{(1)}_{ES}$ in B(a)-B(c), it is evident that $C^{(2)}_{ES}$ provides better insight 
about CHA. Hence, we have presented $C_{E_{\rvec}S_{\rvec}}^{(2)},~C_{E_{\pvec}S_{\pvec}}^{(2)},~C_{E_{t}S_{t}}^{(2)}$ at some
selected $r_c$ (not same for all states) in Table~I, for $1s, 2s, 2p, 3d$, while the remaining states ($3s, 3p, 4f, 5g$) are 
offered in Table~S1. These results corroborate the conclusions drawn from Figs.~1 and S1. None of these could be directly 
compared with literature data, as no such works exist, to the best of our knowledge. 

\begin{figure}                         
\begin{minipage}[c]{0.30\textwidth}\centering
\includegraphics[scale=0.45]{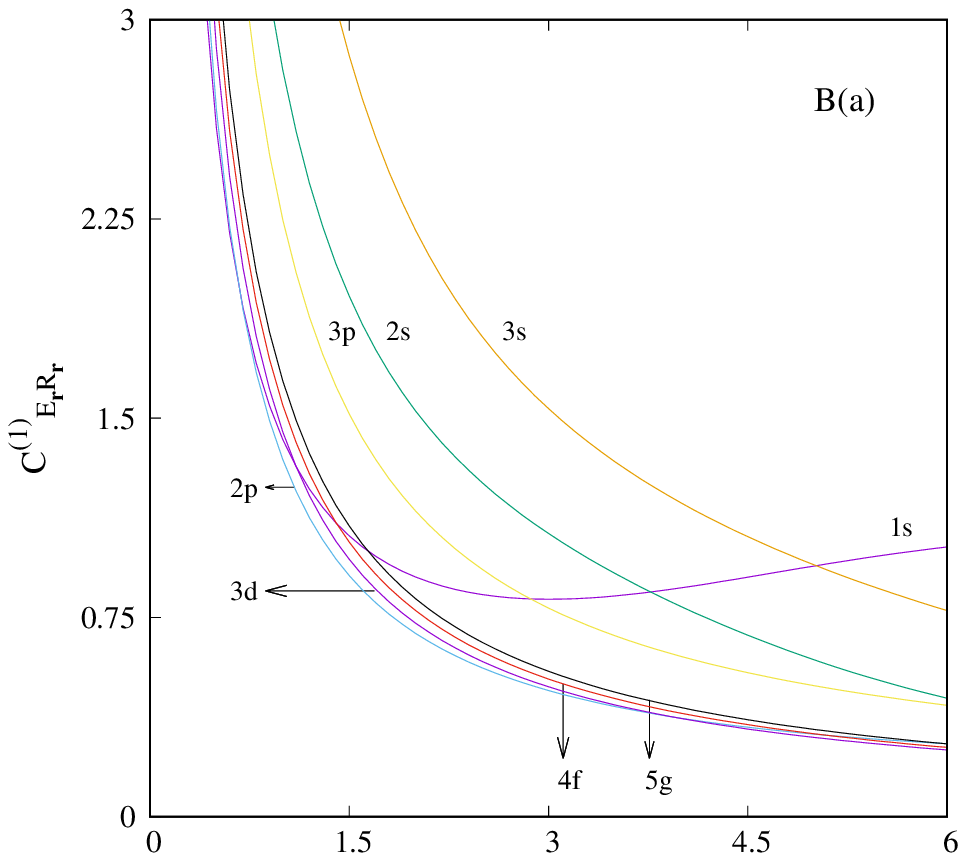}
\end{minipage}%
\vspace{1mm}
\begin{minipage}[c]{0.30\textwidth}\centering
\includegraphics[scale=0.45]{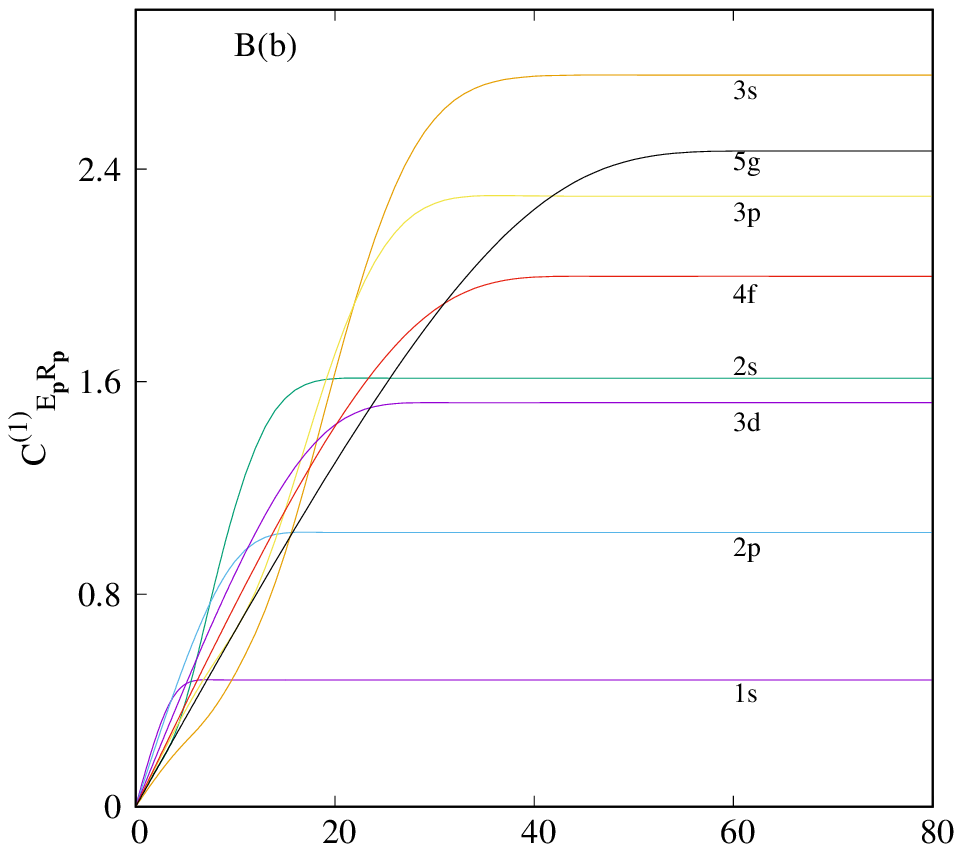}
\end{minipage}%
\vspace{1mm}
\hspace{0.2in}
\begin{minipage}[c]{0.32\textwidth}\centering
\includegraphics[scale=0.46]{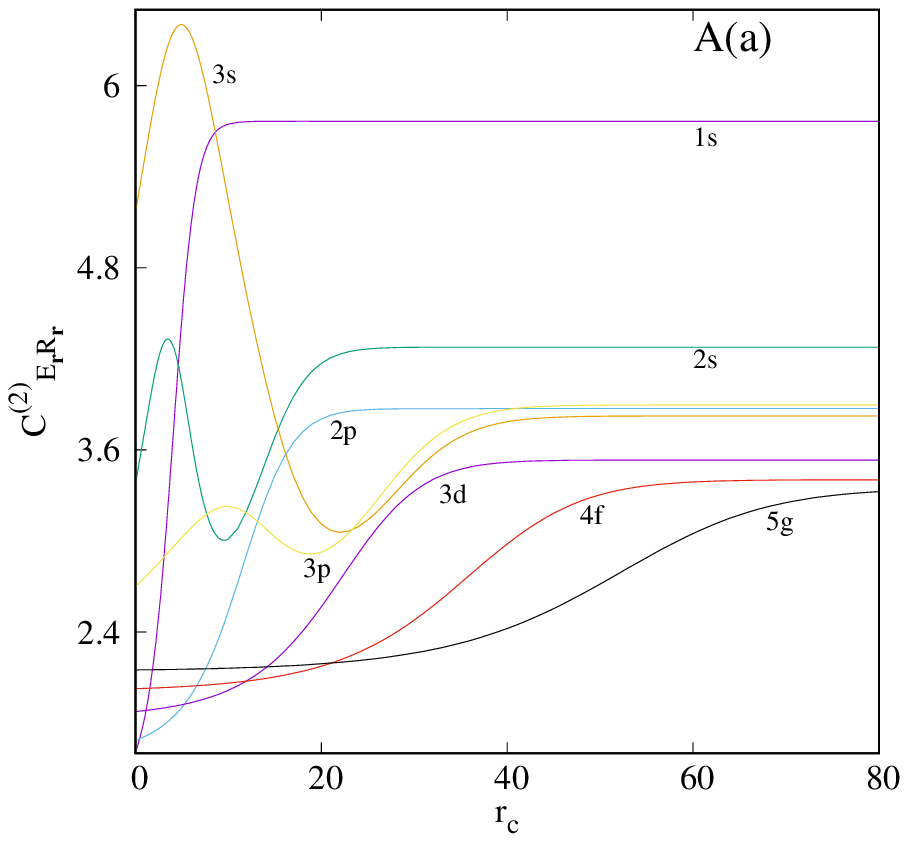}
\end{minipage}%
\begin{minipage}[c]{0.32\textwidth}\centering
\includegraphics[scale=0.46]{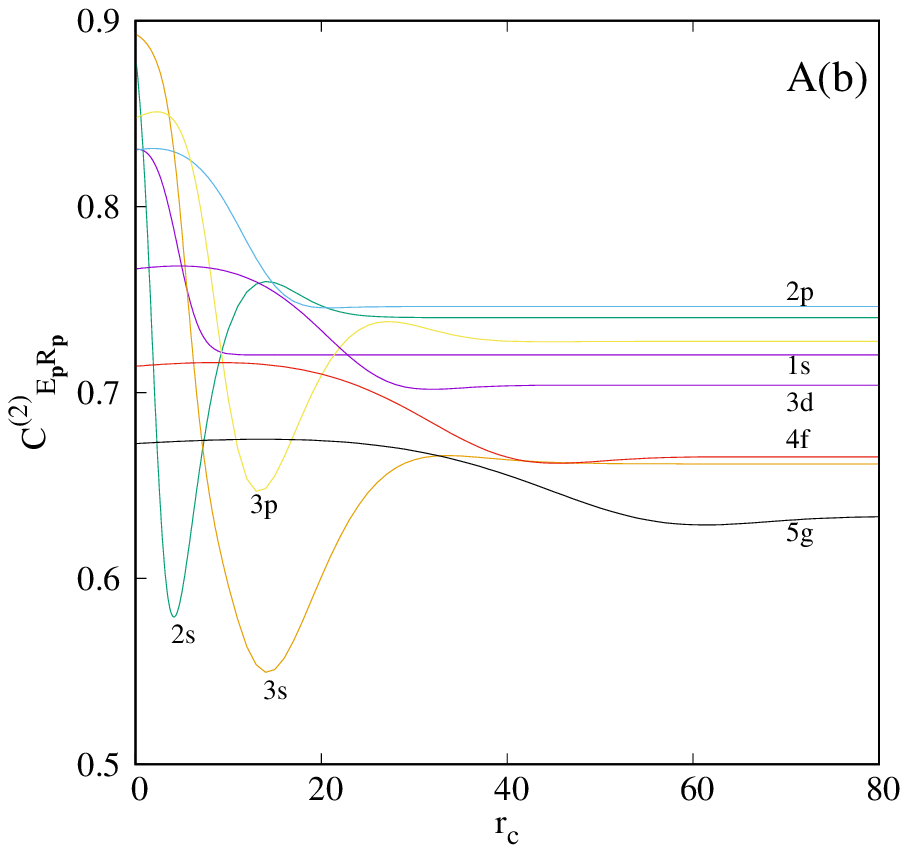}
\end{minipage}%
\caption{Changes in $C_{E_{\rvec}R_{\rvec}}^{(2)},~C_{E_{\pvec}R_{\pvec}}^{(2)}$ (bottom row A) and 
$C_{E_{\rvec}R_{\rvec}}^{(1)},~C_{E_{\pvec}R_{\pvec}}^{(1)}$ (top row B) in CHA with $r_c$ for $1s$-$3d, 4f$ and $5g$ 
states. For more details, see text.}
\end{figure}

\begingroup            
\squeezetable
\begin{table}
\caption{$C_{E_{\rvec}R_{\rvec}}^{(2)},~C_{E_{\pvec}R_{\pvec}}^{(2)}$ and $C_{E_{t}R_{t}}^{(2)}$ for $1s,~2s,~2p,~3d$ 
states in CHA at various $r_c$.}
\centering
\begin{ruledtabular}
\begin{tabular}{llll|llll}
$r_c$  &  $C_{E_{\rvec}R_{\rvec}}^{(2)}$ & $C_{E_{\pvec}R_{\pvec}}^{(2)}$  & $C_{E_{t}R_{t}}^{(2)}$ & $r_c$ 
       & $C_{E_{\rvec}R_{\rvec}}^{(2)}$ 
& $C_{E_{\pvec}R_{\pvec}}^{(2)}$  & $C_{E_{t}R_{t}}^{(2)}$ \vspace{1mm}  \\\hline
\multicolumn{4}{c}{$1s$} \vline &  \multicolumn{4}{c}{\hspace{-5mm}$2s$}    \\
\hline
 0.1    &    1.62388648   &  0.83076854  &  1.34907381   & 0.1     & 3.40812187  &   0.876978     &  2.988848         \\
 1.5    &    2.00911324   &  0.82750748  &  1.66255624  &  2.0     & 4.07095921  &   0.707115     &  2.878639          \\
 3.3    &    3.06621535   &  0.80540213  &  2.46953640  &  3.5     & 4.32946637  &   0.589721     &  2.553178          \\
 4.5    &    4.06762769   &  0.77911543  &  3.16915152  &  6.5     & 3.49223860  &   0.644208     &  2.249730          \\
 6.0    &    5.08995204   &  0.74682789  &  3.80131819  &  9.6     & 3.00374145  &   0.727635     &  2.185627          \\
18.0    &    5.76509705   &  0.72020586  &  4.15205673  & 30.0     & 4.27524770  &   0.740362     &  3.165232          \\
25.0    &    5.76468568   &  0.72020460  &  4.15175319  & 42.0     & 4.27629214  &   0.740292     &  3.165706          \\
30.0    &    5.76468568   &  0.72020460  &  4.15175319  & 50.0     & 4.27629299  &   0.740292     &  3.165706          \\
\hline
\multicolumn{4}{c}{$2p$}  \vline  &   \multicolumn{4}{c}{\hspace{-5mm}$3d$}    \\
\hline
 0.1    &  1.68688780      & 0.830403   & 1.400796 &  0.1      & 1.875596  &     0.766519   & 1.437680    \\
 6.2    &  2.00463718      & 0.823814   & 1.651448 &  9.5      & 2.002374  &     0.765598   & 1.533015    \\
 9.9    &  2.51770021      & 0.800244   & 2.014775 & 20.0      & 2.565092  &     0.733376   & 1.881177    \\
12.0    &  2.91667206      & 0.781037   & 2.278030 & 25.0      & 3.004262  &     0.711700   & 2.138136    \\
15.0    &  3.43010804      & 0.757407   & 2.597988 & 35.0      & 3.472244  &     0.702409   & 2.438936    \\
40.0    &  3.87153850      & 0.746249   & 2.889133 & 50.0      & 3.531830  &     0.703914   & 2.486106    \\
50.0    &  3.87153994      & 0.746249   & 2.889134 & 70.0      & 3.532549  &     0.703904   & 2.486576    \\
60.0    &  3.87153994      & 0.746249   & 2.889134 & 80.0      & 3.532549  &     0.703904   & 2.486576    \\
\end{tabular}
\end{ruledtabular}
\end{table}
\endgroup 

Similarly, bottom row of Fig.~2 illustrates the behavior of $C_{E_{\rvec}R_{\rvec}}^{(2)},~C_{E_{\pvec}R_{\pvec}}^{(2)}$ with 
changes in $r_{c}$ for the same states of Fig.~1. Panel A(a) shows that, like $C_{E_{\rvec}S_{\rvec}}^{(2)}$, here also
$C_{E_{\rvec}R_{\rvec}}^{(2)}$ for circular states advances to their respective limiting values with growth of $r_c$. 
The nodal states ($2s,3s,3p$) once again display similar pattern as in $C_{E_{\rvec}S_{\rvec}}^{(2)}$. At first they cross 
through a maximum followed by a minimum before eventually coalescing to the respective FHA values. In stronger confinement region, 
($r_c \lessapprox 0.4$), at a certain $r_{c}$, $C_{E_{\rvec}R_{\rvec}}^{(2)}$ enhances with $n$ for the circular states. But 
for higher $r_{c}$ (FHA limit), there is a decrement in the same for these five states with betterment of $n$. From panel~A(b) 
it is vivid that, for all these circular states, $C_{E_{\pvec}S_{\pvec}}^{(2)}$ diminish with $r_c$, then attains a shallow 
minimum and finally stretches to respective FHA value. But for $2s,~3s,~3p$ states, prominent minimum pursued by small maximum 
are observed (before reaching FHA value). Position of these extrema get right shifted with $n$. Moreover, the depth of the 
minimum enhances with rise of $n$ within a fixed $l$. The relevant total measures are again displayed in Fig.~S2 of supplementary
material. For nodeless states, $C_{E_{t}R_{t}}^{(2)}$, in panel A(c), like $C_{E_{t}S_{t}}^{(2)}$ advances to their FHA value with 
progression in $r_c$, while for $2s,~3s,~3p$ it passes through a maximum and a minimum before reaching their the same limit. 
Now panels B(a) and B(b) in top row indicate that, for all these states $C_{E_{\rvec}R_{\rvec}}^{(1)}$ gradually decline and 
$C_{E_{\pvec}R_{\pvec}}^{(1)}$ show opposite trend with advances in $r_c$. However, panel B(c) of Fig.~S2 imprints that, 
$C_{E_{t}R_{t}}^{(1)}$ for circular states improves with $r_c$. But for nodal ($2s,~2p,~3s$) states it reaches the FHA threshold
by passing consecutive maximum and a minimum. This once again suggests that out of $C^{(2)}_{ER}$ and $C^{(1)}_{ER}$, 
the former offers more detailed knowledge about CHA, which justifies the quantities produced in Table~II, namely, 
$C_{E_{\rvec}R_{\rvec}}^{(2)},~C_{E_{\pvec}R_{\pvec}}^{(2)}$ and $C_{E_{t}R_{t}}^{(2)}$. These are given for four states 
($1s, 2s, 2p, 3d$) at eight suitably chosen $r_c$ (not same for all states); Table~S2 presents same for $3s, 3p, 4f, 5g$ states. 
These two Tables~II and S2 complement the inferences drawn from Figs.~2 and S2. As in the previous table, here also no literature
results could be quoted. Additionally, in $r$ and $p$ spaces $C_{E_{\rvec}R_{\rvec}}^{(2)}$ and $C_{E_{\pvec}R_{\pvec}}^{(2)}$ 
exhibit opposite behavior but for $C_{E_{\rvec}S_{\rvec}}^{(2)}$ and $C_{E_{\pvec}S_{\pvec}}^{(2)}$ an analogous trend is 
observed. Hence, $C_{E_{\rvec}R_{\rvec}}^{(2)}$ and $C_{E_{\pvec}R_{\pvec}}^{(2)}$ turn out to be a relatively better measure 
of complexity.

\begin{figure}                         
\begin{minipage}[c]{0.30\textwidth}\centering
\includegraphics[scale=0.45]{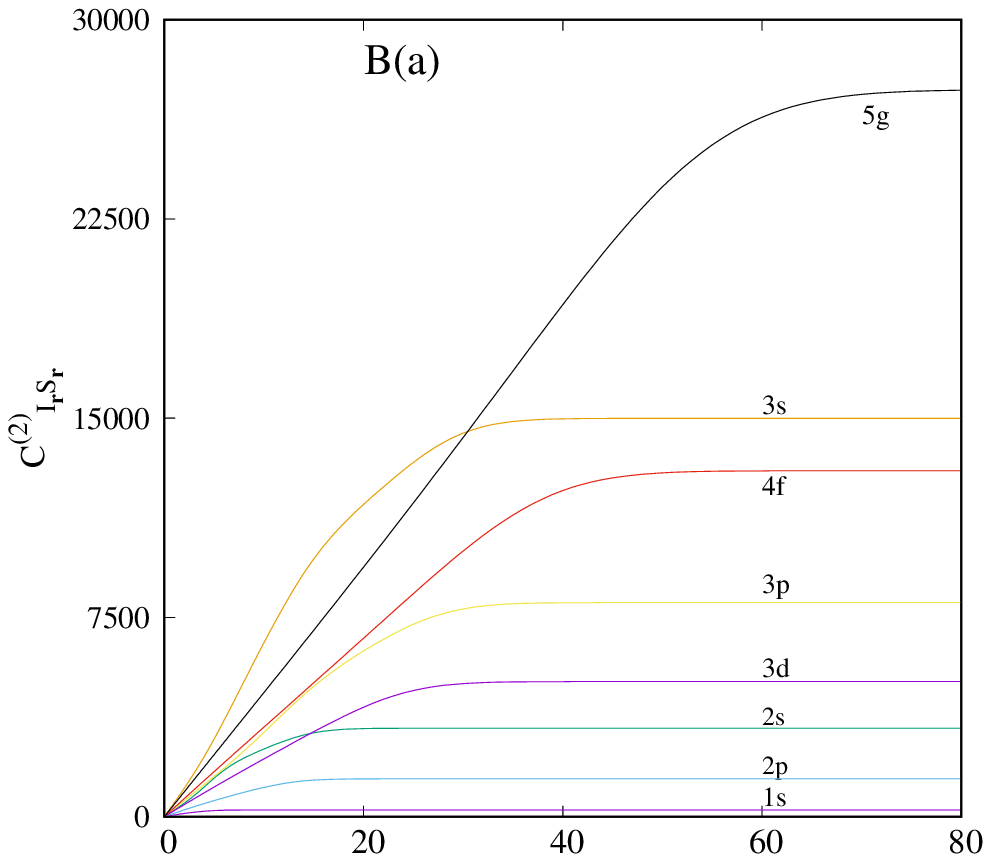}
\end{minipage}%
\vspace{1mm}
\begin{minipage}[c]{0.30\textwidth}\centering
\includegraphics[scale=0.45]{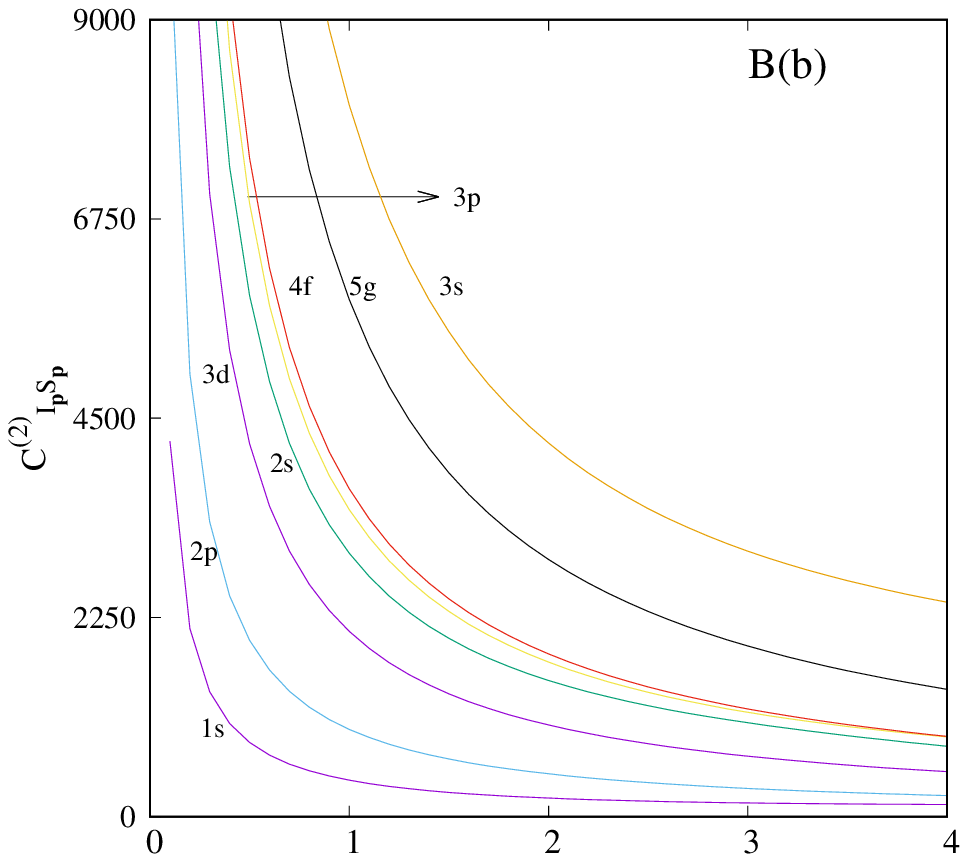}
\end{minipage}%
\vspace{1mm}
\hspace{0.2in}
\begin{minipage}[c]{0.32\textwidth}\centering
\includegraphics[scale=0.46]{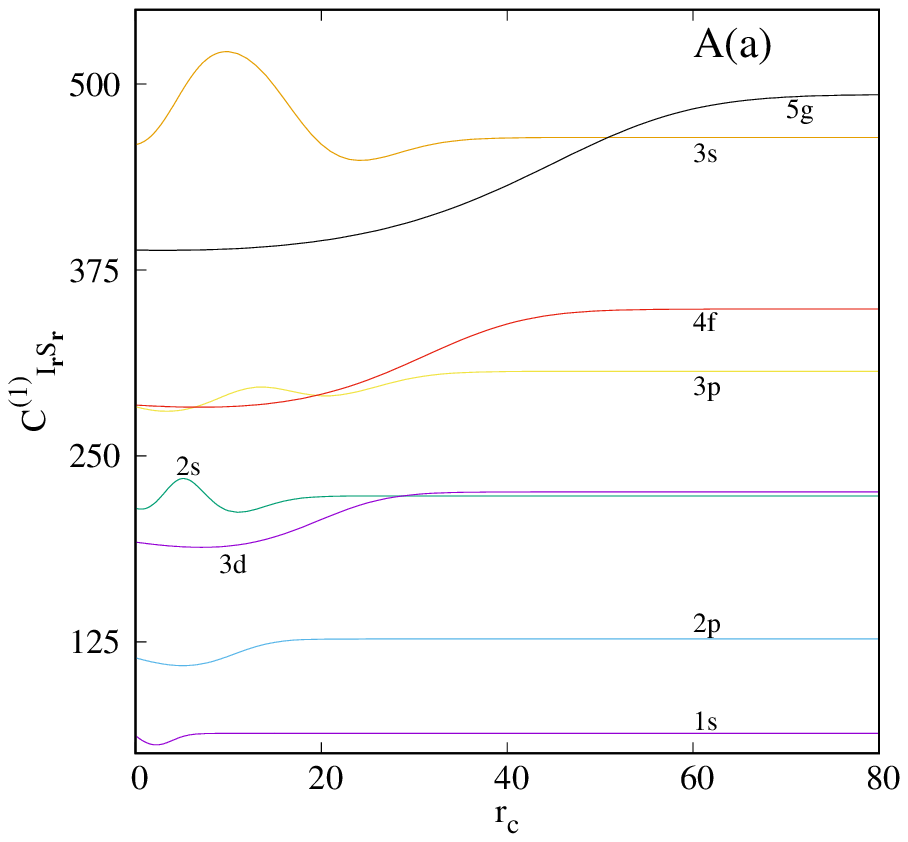}
\end{minipage}%
\begin{minipage}[c]{0.32\textwidth}\centering
\includegraphics[scale=0.46]{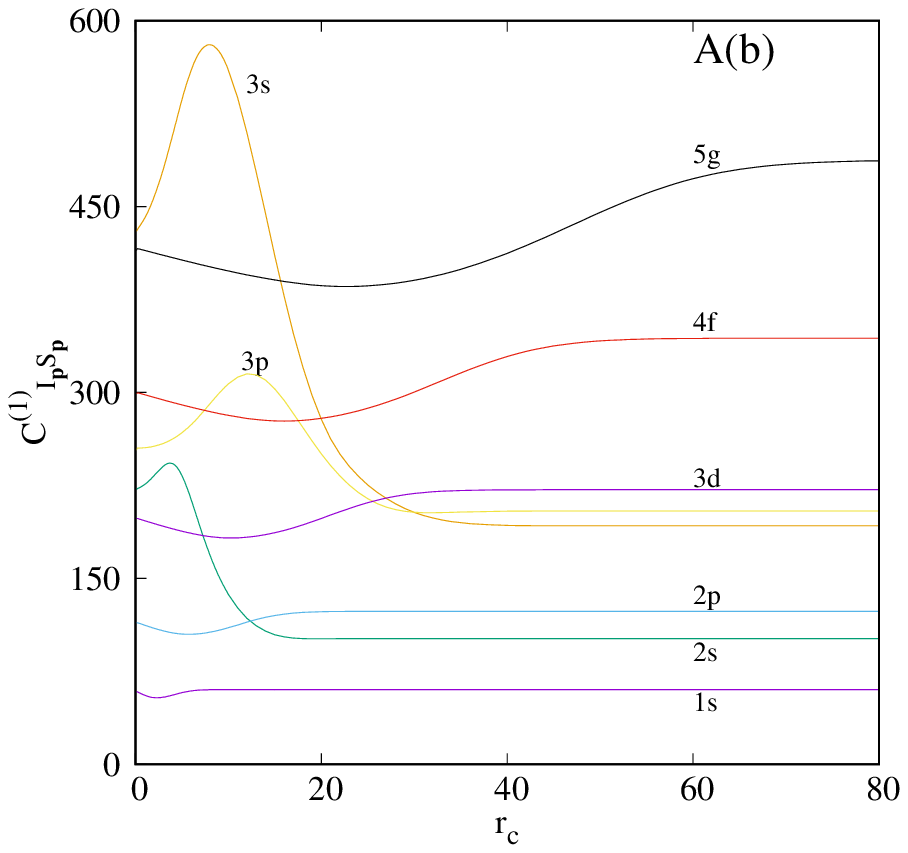}
\end{minipage}%
\caption{Variation of $C_{I_{\rvec}S_{\rvec}}^{(1)},~C_{I_{\pvec}S_{\pvec}}^{(1)}$ (bottom row A)  
$C_{I_{\rvec}S_{\rvec}}^{(2)},~C_{I_{\pvec}S_{\pvec}}^{(2)}$ (top row B)  
in CHA with $r_c$ for $1s$-$3d,4f$ and $5g$ states. Consult text for more details.} 
\end{figure}

Next in Fig.~3 the two rows incorporating panels \{A(a), A(b)\}, \{B(a), B(b)\} portray the variation of 
$\{C_{I_{\rvec}S_{\rvec}}^{(1)},~C_{I_{\pvec}S_{\pvec}}^{(1)}\}$ and 
$\{C_{I_{\rvec}S_{\rvec}}^{(2)},~C_{I_{\pvec}S_{\pvec}}^{(2)}\}$ with changes in $r_{c}$. The bottom two panels as well as 
A(c) of Fig.~S3 illustrate that, for circular states, $C_{I_{\rvec}S_{\rvec}}^{(1)},~C_{I_{\pvec}S_{\pvec}}^{(1)}$ and 
$C_{I_{t}S_{t}}^{(1)}$ initially fall with growth of $r_c$, then attain a minimum and finally converge to respective FHA values. 
But, for the other three states ($2s,~3s,~3p$), $C_{I_{\rvec}S_{\rvec}}^{(1)}$ grows to a maximum then falls down to a minimum 
with advance in $r_c$ and eventually join the FHA result. Whereas, $C_{I_{\pvec}S_{\pvec}}^{(1)}$ and $~C_{I_{t}S_{t}}^{(1)}$, for 
these three non-circular states only pass through a maximum before reaching their borderline values. On the other hand, top 
panels B(a) and B(b) portray that, for all eight states considered, $C_{I_{\rvec}S_{\rvec}}^{(2)}$ and 
$C_{I_{\pvec}S_{\pvec}}^{(2)}$ consistently progress and reduce respectively with increase in $r_c$. However, panel B(c) in 
Fig.~S3 shows that, for circular states $C_{I_{t}S_{t}}^{(2)}$ enhances with development of $r_c$ and for states with nodes, 
it passes through a maximum before merging to FHA result. A careful study of Figs.~3 and S3 reveals another interesting feature
that, in case of CHA, $C_{I_{\rvec}S_{\rvec}}^{(1)},~C_{I_{\pvec}S_{\pvec}}^{(1)},~C_{I_{t}S_{t}}^{(1)}$ provides a more  
detailed account than $C_{I_{\rvec}S_{\rvec}}^{(2)},~C_{I_{\pvec}S_{\pvec}}^{(2)},~C_{I_{t}S_{t}}^{(2)}$. Thus, to get a 
quantitative idea, $C_{I_{\rvec}S_{\rvec}}^{(1)},~C_{I_{\pvec}S_{\pvec}}^{(1)},~C_{I_{t}S_{t}}^{(1)}$ values at some selected 
$r_c$'s are given in Tables~III ($1s, 2s, 2p, 3d$) and S3 ($3s, 3p, 4f, 5g$). Again no results are available in literature except 
the lone ground state for $C_{IS}^{(1)}$ at few $r_c$ values, which are duly quoted for comparison (for $r_c=$0.1, 1, 30, 40). 
Current results are in good agreement with the reported one and may be useful for future references. 

\begingroup         
\squeezetable
\begin{table}
\caption{$C_{I_{\rvec}S_{\rvec}}^{(1)},~C_{I_{\pvec}S_{\pvec}}^{(1)}$ and $C_{I_{t}S_{t}}^{(1)}$ for $1s,~2s,~2p,~3d$ states in 
CHA at selected $r_c$.}
\centering
\begin{ruledtabular}
\begin{tabular}{llll|llll}
$r_c$  &  $C_{I_{\rvec}S_{\rvec}}^{(1)}$ & $C_{I_{\pvec}S_{\pvec}}^{(1)}$  & $C_{I_{t}S_{t}}^{(1)}$ & $r_c$ & 
$C_{I_{\rvec}S_{\rvec}}^{(1)}$ & $C_{I_{\pvec}S_{\pvec}}^{(1)}$ 
 & $C_{I_{t}S_{t}}^{(1)}$ \vspace{1mm}  \\ 
\hline
\multicolumn{4}{c}{$1s^{\dag}$} \vline &  \multicolumn{4}{c}{\hspace{-5mm}$2s$}    \\
\hline

  0.1\footnotemark[1]    &    61.445391   & 58.4510   &  3591.5482  &	 0.1      &   214.635698    &   221.5739     &  47557.6874 \\
  1.0\footnotemark[2]    &    57.747939   & 55.7023   &  3216.6965  &	 0.7      &   214.216370    &   224.3204     &  48053.1193 \\
  2.3    &    55.670758   & 53.7138   &  2990.2906  &	 3.0      &   224.290385    &   240.5022     &  53942.3512          \\
  4.0    &    58.707169   & 55.8278   &  3277.4955  &	 5.2      &   234.766329    &   229.7707     &  53942.4448          \\
  4.7    &    60.463080   & 57.2340   &  3460.5492  &	11.0      &   212.022314    &   126.2148     &  26760.3588          \\
  6.5    &    62.907313   & 59.6342   &  3751.4320  &	17.0      &   220.839663    &   101.9061     &  22504.9152          \\
 30.0\footnotemark[3]    &    63.398969   & 60.3065   &  3823.3703  &	43.0      &   223.025525    &   101.3888     &  22612.3074          \\
 40.0\footnotemark[4]    &    63.398969   & 60.3065   &  3823.3703  &	50.0      &   223.025525    &   101.3889     &  22612.3209          \\
\hline
\multicolumn{4}{c}{$2p$}  \vline  &   \multicolumn{4}{c}{\hspace{-5mm}$3d$}    \\
\hline
   0.1    &  114.078664      & 114.3281   & 13042.3998 &   0.1      & 191.947614  &  198.3921      &  38080.8988        \\
   2.5    &  110.537817      & 108.7010   & 12015.5737 &   3.0      & 189.827438  &  192.0738      &  36460.8929        \\
   5.1    &  108.960438      & 105.0430   & 11445.5361 &   7.2      & 188.511787  &  184.6662      &  34811.7628        \\
   7.1    &  110.071219      & 105.5415   & 11617.0842 &  12.0      & 191.035748  &  183.2506      &  35007.4345        \\
   9.5    &  114.166616      & 109.1567   & 12462.0605 &  17.0      & 199.745109  &  190.8078      &  38112.9401        \\
  15.0    &  124.642286      & 120.1211   & 14972.1809 &  21.0      & 209.748916  &  201.1179      &  42184.2742        \\
  34.0    &  126.882910      & 123.4402   & 15662.4635 &  56.0      & 225.764533  &  221.5783      &  50024.5236        \\
  50.0    &  126.882916      & 123.4403   & 15662.4772 &  70.0      & 225.764533  &  221.5783      &  50024.5236        \\
 \end{tabular}
 \end{ruledtabular}
\begin{tabbing}
$^{\mathrm{a}}${Reference result~\cite{aquino13}: $C_{I_{\rvec}S_{\rvec}}^{(1)}=61.4476,~C_{I_{\pvec}S_{\pvec}}^{(1)}=58.9580,~C_{I_{t}S_{t}}^{(1)}=3622.8276$} \\    
$^{\mathrm{b}}${Reference result~\cite{aquino13}: $C_{I_{\rvec}S_{\rvec}}^{(1)}=57.7561,~C_{I_{\pvec}S_{\pvec}}^{(1)}=55.6956,~C_{I_{t}S_{t}}^{(1)}=3216.7606$} \\                  
$^{\mathrm{c}}${Reference result~\cite{aquino13}: $C_{I_{\rvec}S_{\rvec}}^{(1)}=63.4008,~C_{I_{\pvec}S_{\pvec}}^{(1)}=60.3087,~C_{I_{t}S_{t}}^{(1)}=3823.6198$}    \\        
$^{\mathrm{d}}${Reference result~\cite{aquino13}: $C_{I_{\rvec}S_{\rvec}}^{(1)}=63.4008,~C_{I_{\pvec}S_{\pvec}}^{(1)}=60.3087,~C_{I_{t}S_{t}}^{(1)}=3823.6198$}  \\              
$^{\dag}${Reference values are multiplied with a 8$\pi^{2}e$ factor in both $r$ and $p$ space.}  
\end{tabbing}
\end{table}
\endgroup

Finally, in Fig.~4 the two lower (A(a),A(b)) and upper (B(a)-B(b)) panels depict the alteration of our last complexity 
measure, \emph{viz.,} $C_{I_{\rvec}R_{\rvec}}^{(1)},~C_{I_{\pvec}R_{\pvec}}^{(1)}$ and 
$C_{I_{\rvec}R_{\rvec}}^{(2)},~C_{I_{\pvec}R_{\pvec}}^{(2)}$ with variation in $r_{c}$. Here, again the bottom row as well as
panel A(c) of Fig.~S4 shows that, for circular states, $C_{I_{\rvec}R_{\rvec}}^{(1)},~C_{I_{\pvec}R_{\pvec}}^{(1)}$ and 
$C_{I_{t}R_{t}}^{(1)}$ reduce to attain a minimum with rise of $r_c$ and finally assume the fate of FHA. But for three remaining 
nodal states, $C_{I_{\rvec}R_{\rvec}}^{(1)}$ progress via a maximum and minimum successively with growth in $r_c$; after that 
they approach the FHA result. However, $C_{I_{\pvec}R_{\pvec}}^{(1)}$ and $~C_{I_{t}R_{t}}^{(1)}$ for the aforesaid states 
with nodes, climb a maximum and fall down asymptotically to a constant FHA value. Besides these, panels B(a) and B(b) portray 
that, for all these reported eight states concerned, $C_{I_{\rvec}R_{\rvec}}^{(2)}$ and $C_{I_{\pvec}R_{\pvec}}^{(2)}$ 
rise and fall respectively with growth of $r_c$. However, panel B(c) in Fig.~S4 shows that, for circular states 
$C_{I_{t}R_{t}}^{(2)}$ improve with elevation of $r_c$ and for $2s,3s,3p$ states it proceeds through a maximum before reaching 
the limiting value at FHA. A closer investigation of Figs.~4 and S4 conveys that, 
$C_{I_{\rvec}R_{\rvec}}^{(1)},~C_{I_{\pvec}R_{\pvec}}^{(1)},~C_{I_{t}R_{t}}^{(1)}$ scale the system of this orientation better
than $C_{I_{\rvec}R_{\rvec}}^{(2)},~C_{I_{\pvec}R_{\pvec}}^{(2)},~C_{I_{t}R_{t}}^{(2)}$. Hence, to conclude, the former three 
measures are offered in Tables~IV and S4, at some appropriately chosen $r_c$. No comparison could be made due to lack of any 
literature data and hopefully these would be useful in future. 

\begin{figure}                         
\begin{minipage}[c]{0.30\textwidth}\centering
\includegraphics[scale=0.45]{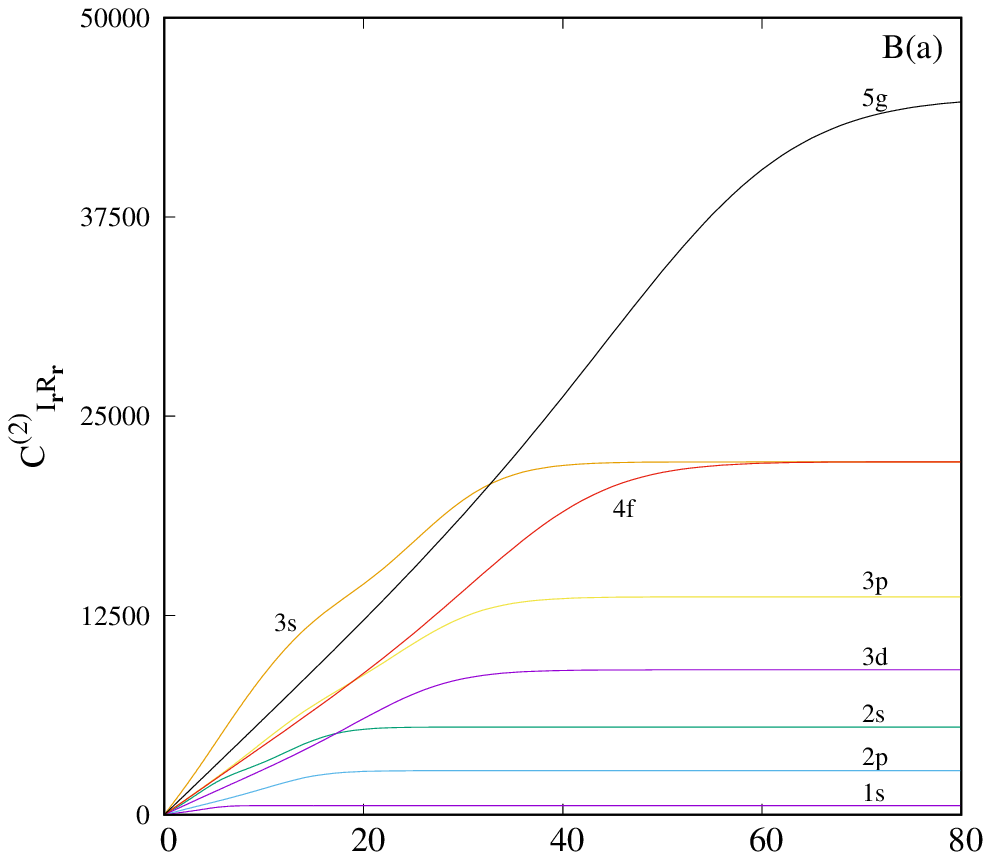}
\end{minipage}%
\vspace{1mm}
\begin{minipage}[c]{0.30\textwidth}\centering
\includegraphics[scale=0.45]{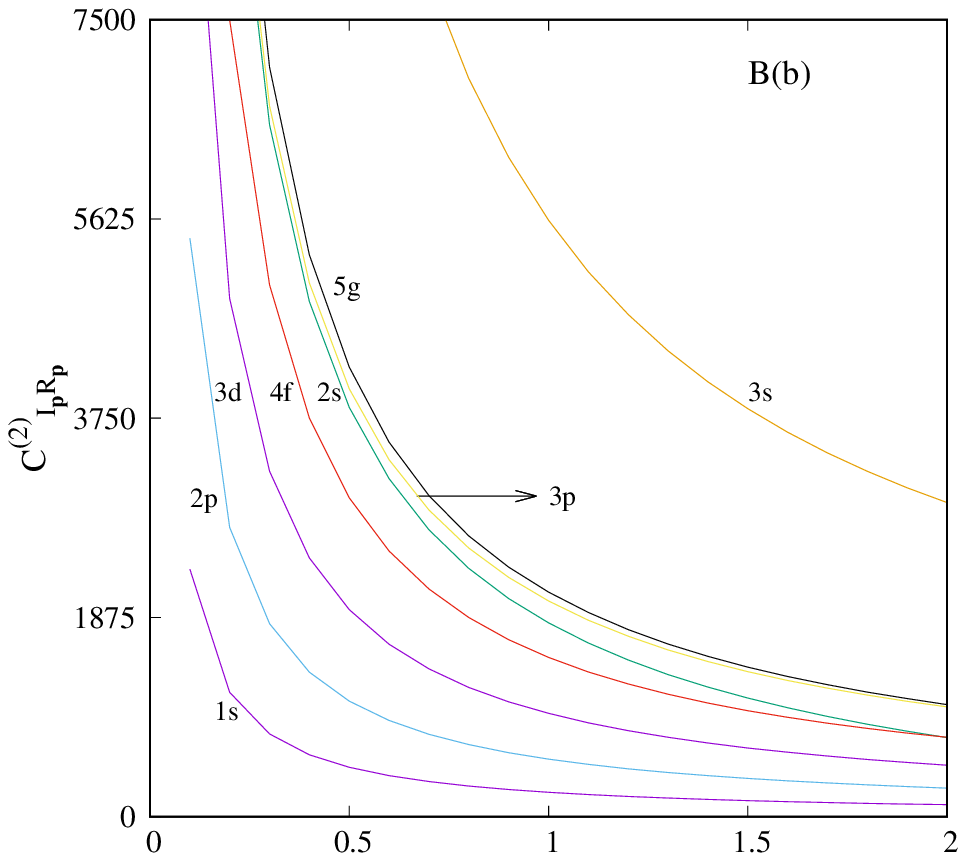}
\end{minipage}%
\vspace{1mm}
\hspace{0.2in}
\begin{minipage}[c]{0.32\textwidth}\centering
\includegraphics[scale=0.46]{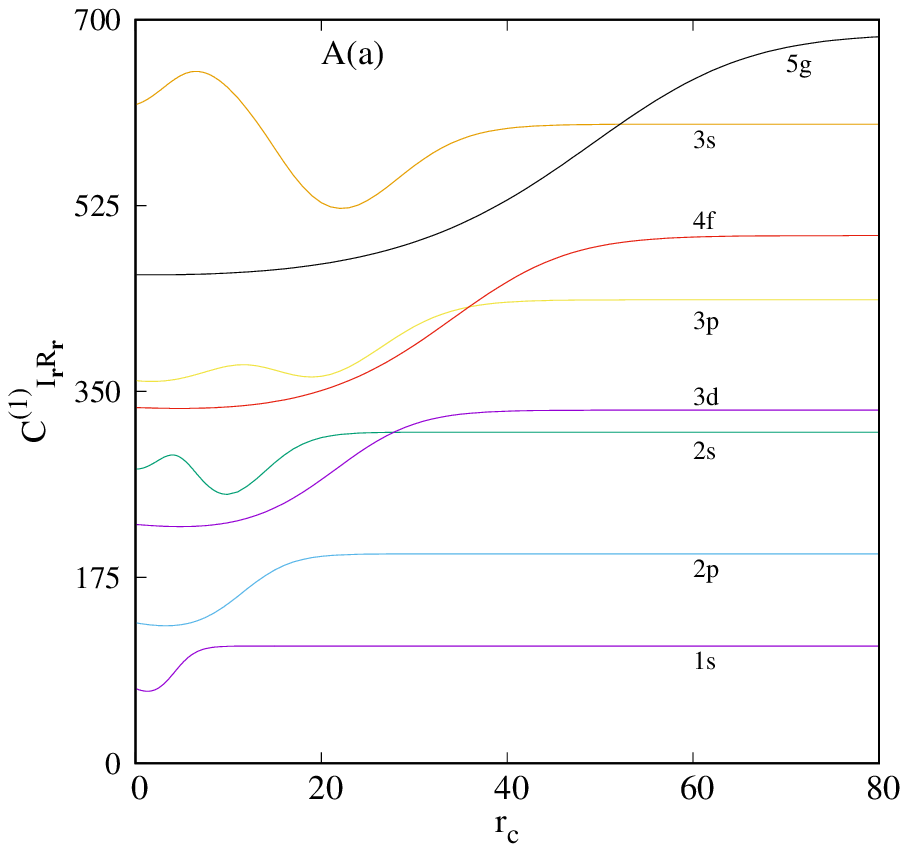}
\end{minipage}%
\begin{minipage}[c]{0.32\textwidth}\centering
\includegraphics[scale=0.46]{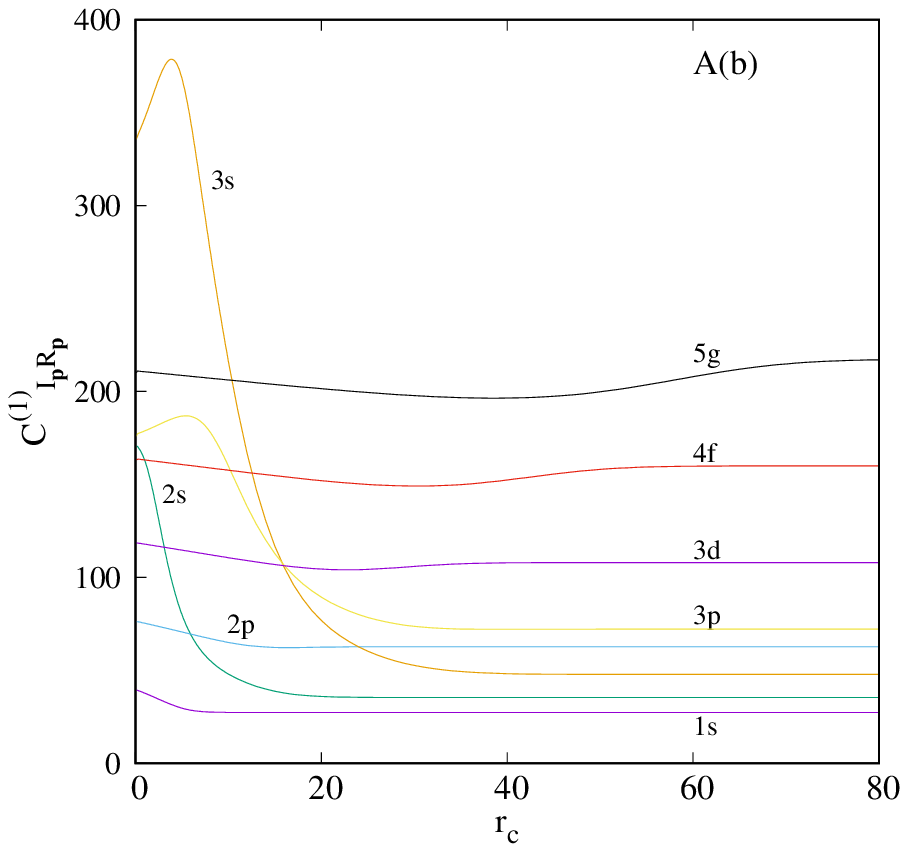}
\end{minipage}%
\caption{Plots of $C_{I_{\rvec}R_{\rvec}}^{(1)},~C_{I_{\pvec}R_{\pvec}}^{(1)}$ (bottom row A) and 
$C_{I_{\rvec}R_{\rvec}}^{(2)},~C_{I_{\pvec}R_{\pvec}}^{(2)}$ (top row B) in CHA with $r_c$ for $1s$-$3d, 4f$ and $5g$
states. For further details, see text.}
\end{figure}

\section{Future and outlook}
Various complexity measures like $C_{ES},~C_{IS},~C_{ER},~C_{IR}$ are explored for low-lying states of CHA in both $r$ and $p$ 
space, keeping $m$ fixed at zero. We have pursued our calculation using both global quantity ($E$) and local quantity ($I$) as 
a measure of $order$ in a system. Except for some results of $C_{IS}$ in ground state, all these quantities are reported here 
for the first time. It is found that, $C^{(2)}_{ES},~C^{(2)}_{ER}$ offer more detailed explanation than 
$C^{(1)}_{ES},~C^{(1)}_{ER}$ about the system. On the contrary, $C^{(1)}_{IS},~C^{(1)}_{IR}$ interpret the behavior of CHA more 
efficiently than that of $C^{(2)}_{IS},~C^{(2)}_{IR}$. Hence, depending upon the nature of complexity measures, it is necessary 
to determine the appropriate value of $b$.  Accurate results for $C^{(2)}_{ES},C^{(2)}_{ER},C^{(1)}_{IS},C^{(2)}_{IR}$ 
(radial \emph{plus} angular) are provided for $1s$-$3d,~4f$ and $5g$ states of CHA, most of them for the first time. Further, 
an investigation of all these quantities in the realm of Rydberg states under different kinds of confined environment, as well as
the correlation between complexity and periodicity in many-electron atomic systems may be worthwhile pursuing. 

\begingroup           
\squeezetable
\begin{table}
\caption{$C_{I_{\rvec}R_{\rvec}}^{(1)},~C_{I_{\pvec}R_{\pvec}}^{(1)}$ and $C_{I_{t}R_{t}}^{(1)}$ for $1s,~2s,~2p,~3d$ states in 
CHA at some selected $r_c$.}
\centering
\begin{ruledtabular}
\begin{tabular}{llll|llll}
$r_c$  &  $C_{I_{\rvec}R_{\rvec}}^{(1)}$ & $C_{I_{\pvec}R_{\pvec}}^{(1)}$  & $C_{I_{t}R_{t}}^{(1)}$ & $r_c$ & $C_{I_{\rvec}R_{\rvec}}^{(1)}$ & $C_{I_{\pvec}R_{\pvec}}^{(1)}$  & $C_{I_{t}R_{t}}^{(1)}$ \vspace{1mm}  \\ 
\hline
\multicolumn{4}{c}{$1s$} \vline &  \multicolumn{4}{c}{\hspace{-5mm}$2s$}    \\
\hline

  0.1   &   70.18835424     &   39.206889  &  2751.867082   &   0.1     &  276.919929   &    170.569679    & 47234.143661      \\
  1.0   &   67.97441474     &   37.745459  &  2565.725509   &   2.0     &  282.717996   &    143.626221    & 40605.717425      \\
  2.5   &   70.98469720     &   34.502985  &  2449.183973   &   4.0     &  290.232344   &     96.448638    & 27992.514409      \\
  5.0   &   93.90104729     &   29.677676  &  2786.764914   &   7.0     &  268.480761   &     61.038709    & 16387.719163      \\
  5.8   &  100.48737687     &   28.715280  &  2885.523167   &   9.8     &  253.236641   &     48.608701    & 12309.504305      \\
  6.8   &  105.75702471     &   27.975405  &  2958.595632   &  18.0     &  300.561968   &     36.584635    & 10995.950138      \\
 23.0   &  110.34101779     &   27.379301  &  3021.060018   &  50.0     &  311.686645   &     35.411689    & 11037.350749      \\
 30.0   &  110.34101779     &   27.379301  &  3021.060018   &  70.0     &  311.686648   &     35.411668    & 11037.344253      \\
\hline
\multicolumn{4}{c}{$2p$}  \vline  &   \multicolumn{4}{c}{\hspace{-5mm}$3d$}    \\
\hline
  0.1   &    132.174030     & 76.150268  & 10065.087867 &   0.1	  &  224.853345  & 118.438787	 &  26631.357655           \\
  3.4   &    129.499950     & 72.573286  &  9398.237011 &  16.0   &  244.855787  & 106.300730    &  26028.349158           \\
  5.8   &    132.007059     & 69.694271  &  9200.135775 &  20.0   &  266.977357  & 104.494337    &  27897.622145           \\
  8.2   &    140.266419     & 66.892350  &  9382.750441 &  23.0   &  286.056400  & 104.096971    &  29777.604868           \\
 10.0   &    150.439765     & 65.033889  &  9783.683044 &  26.0   &  303.369560  & 104.554144    &  31718.544807           \\
 13.0   &    170.727938     & 62.902501  & 10739.214335 &  56.0   &  332.420701  & 107.955748    &  35886.725520           \\
 60.0   &    197.127576     & 62.701899  & 12360.273469 &  86.0   &  332.423192  & 107.955697    &  35886.977628           \\
 70.0   &    197.127576     & 62.701899  & 12360.273469 &  90.0   &  332.423192  & 107.955697    &  35886.977628           \\
\end{tabular}
\end{ruledtabular}
\end{table}
\endgroup

\section{Acknowledgement}
Financial support from DST SERB, New Delhi, India (sanction order: EMR/2014/000838) is gratefully acknowledged. SM is obliged to 
IISER-K for supporting her JRF. NM thanks DST SERB, New Delhi, India, for a National-post-doctoral fellowship 
(sanction order: PDF/2016/000014/CS).

\end{document}